\documentclass[aps,twocolumn,showpacs,groupedaddress]{revtex4}
\usepackage{amssymb}
\usepackage{graphicx}
\usepackage{amsmath}
\newcommand{\Tr}{\text{Tr}}
\newcommand{\ket}[1]{|#1\rangle}
\newcommand{\bra}[1]{\langle#1|}
\begin{document}
\title{Optimal Lyapunov-based quantum control for quantum systems}
\author{S. C. Hou, M. A. Khan, X. X. Yi,}
\affiliation{School of Physics and Optoelectronic Technology, Dalian
University of Technology, Dalian 116024, China}
\author{Daoyi Dong, Ian R. Petersen}
\affiliation{School of Engineering and Information Technology,
University of New South Wales at the Australian Defence Force
Academy, Canberra, ACT 2600, Australia}
\date{\today}
\begin{abstract}
Quantum Lyapunov control was developed in order to transform  a
quantum system from arbitrary initial states to a target state. The
idea is to find control fields that steer the Lyapunov function to
zero as $t\rightarrow \infty$, meanwhile  the quantum system is
driven to the target state. In order to shorten the time required to
reach the target state, we propose two designs  to optimize Lyapunov
control in this paper. The first design makes the Lyapunov function
decrease as fast as possible with a constraint on the total power of
control fields, and the second design has the same purpose but with
a constraint on each control field. Examples of a three-level system
demonstrate that the evolution time for Lyapunov control can be
significantly shortened, especially when high control fidelity is
required. Besides, this optimal Lyapunov-based quantum control is
robust against uncertainties in the free Hamiltonian and decoherence
in the system compared to conventional Lyapunov control. \textbf{We
apply  our optimal design to cool a nanomechanical resonator, a
shorter cooling time is found with respect to the cooling time by
the conventional Lyapunov design.}
\end{abstract}
\pacs{03.65.Yz, 02.30.Yy, 03.67.-a}\maketitle

\section{introduction}

Quantum control \cite{Alessandro,Dong} has attracted much attention
in recent years and it has found potential applications in many
fields such as quantum information processing, quantum chemistry and
quantum simulation. Among these quantum control problems, state
transfer is a central task. Various methods such as quantum optimal
control and Lie group decompositions have been used to design control
laws to drive quantum systems to target states or to realize some
specific operations \cite{Carlini,Khaneja,Alessandro2,Palao,Schirmer,
Zhou,Law,Liao,Pechen,Dong2,Verstraete}.

Quantum Lyapunov control was proposed in the early 2000s as a good
candidate for state transfer\cite{Grivopoulos,Vettori}. This
strategy has been widely studied recently both in theory and applications
\cite{Mirrahimi,Wang,Wang2,Kuang,Cong,WW,Yi,Wang3,Coron,Beauchard,Yi2},
because it offers a simple and effective way to design control
fields. However, the problem of speeding up Lyapunov control has not
been widely considered to date. Quantum Lyapunov control is practically
employed in an open-loop way without measurement and feedback. Hence,
it is quite natural to shorten the evolution
time so as to overcome the decoherence effect induced by inevitable
interactions with the surrounding environment.

In quantum Lyapunov control, a function $V$, called a Lyapunov
function of quantum states, is specified to design time-varying
control fields. The system converges to the target state given by
$\dot{V}=0$ while $V$ decreases to its minimum. Based on this
concept, we present a scheme to optimize Lyapunov control by using the
following idea: Speed up evolution to the target state by
making $V$ decrease faster.

Generally, the total Hamiltonian in a quantum control system can be
written in two parts $H_0+\sum_{n=1}^{k}f_n(t)H_n$, where $H_0$ is
the free (internal) Hamiltonian and $H_n$ are external control
Hamiltonians with $f_n(t)$ representing the corresponding control
fields. These control fields should be designed to ensure
$\dot{V}\leq0$. In fact, there are many ways to choose $f_n(t)$ to
achieve this goal. In this paper, we consider the question: With
given control Hamiltonians $H_n$, how should we design the shape of the
control fields $f_n(t)$ in order to make $V$ decrease fastest.

It is shown that one can enable $V$ to decrease faster simply  by
enhancing the strength of the control fields. However, strong control
fields may not be feasible and always bring unwanted results, for
example a large energy (power) cost, invalidation of mathematical
approximation and treatment for the system. In view of these
factors, we propose two designs for the control fields. One is under the
constraint that the power-type quantity $W=\sum_{n=1}^k f_n(t)^2$ is
bounded. The other is under the constraint that the strength of each
field is bounded. These designs for control fields make $V$ decrease
as fast as possible within given limitations. The second design
has the simple form of ``bang-bang" control which is easy to be
implemented in an experiment. We also illustrate our control method
with a three-level system. The results suggest that the evolution
time is significantly reduced compared with the conventional  method.

The paper is organized as follows: In Sec.\rm{II}, we  derive
control fields for a general Lyapunov function $V=\Tr(P\rho)$ under
the two aforementioned constraints. In Sec.\rm{III}, we simulate our
field designs and compare them with the conventional  method. The
robustness of the optimal Lyapunov designs is analyzed in
Sec.\rm{IV}.  Finally, we summarize our work in Sec. \rm{V}.

\section{Design of Control fields}
In quantum Lyapunov control, the system is  steered from an initial
state to a target state by control fields designed using Lyapunov
function $V$. The goal of this paper is to obtain optimized control
fields that make the time derivative of the Lyapunov function
$|\dot{V}|$ largest so as  to speed up the evolution to the target
state. We start from a closed quantum system described by the
Liouville equation
\begin{eqnarray}
\frac{d\rho}{dt}=-i[H_0+H_c(t),\rho], \label{eqn:Le}
\end{eqnarray}
where $H_0$ is the free Hamiltonian for the controlled  system and
$H_c(t)$ is a time-dependent Hamiltonian representing coupling to
external control fields which is called the control Hamiltonian. We
have set $\hbar=1$ and assume that the system is controllable. In
Lyapunov control, the solution of Eq.(\ref{eqn:Le}) converges to the
minimum of $V(\rho)$. Meanwhile, the state converges to a set of
states   characterized by the La Salle's invariance principle
\cite{Alessandro}. The control Hamiltonian $H_c(t)$ can be written
in the form,
\begin{equation}
    H_c(t)=\sum_{n=1}^{k}f_n(t)H_n
\label{eqn:CH}
\end{equation}
where $H_n$ $(n=1,...,k)$ are time-independent  Hermitian operators
corresponding to different types of external control and $f_n(t)$
are time-varying real functions, usually representing
electro-magnetic fields. $k$ is a positive integer.

In this paper we consider the following form of Lyapunov function
\begin{eqnarray}
V=\Tr(P\rho), \label{eqn:Lya}
\end{eqnarray}
where $P$ is a Hermitian operator and assumed to be positive
semi-definite in order to satisfy the standard requirement for a
Lyapunov function, $V\geq0$ \cite{Alessandro}. Also, some other forms
of Lyapunov function can be described by Eq.(\ref{eqn:Lya}), such as
that based on the Hilbert Schmidt distance \cite{Alessandro,Kuang}.

The time derivative of $V$ needs to be calculated to design the control fields,
\begin{eqnarray}
\begin{split}
\dot{V}=\Tr(-i P[H_0+\sum_{n=1}^{k}f_n(t)H_n,\rho])
\quad\quad\quad\quad\quad\\
=\Tr(-i\rho[P,H_0])+\sum_{n=1}^kf_n(t)\Tr(-i\rho[P,H_n])\\
=\sum_{n=1}^kf_n(t)T_n
\quad\quad\quad\quad\quad\quad\quad\quad\quad\quad\quad\quad\quad
\end{split}
\label{eqn:dV}
\end{eqnarray}
where $T_n=\Tr(-i\rho[P,H_n])$ is a real function  of $\rho$, $H_n$
and $P$. We have used the assumption that $[P,H_0]=0$, which can be
achieved by constructing $P$ using the eigenvectors of $H_0$.

The Lyapunov control strategy requires $\dot{V}\leq0$. There are
many ways to design $f_n(t)$ to satisfy this
requirement. A simple and conventional way is to let
$f_n(t)=-KT_n$ with $K>0$ so that
\begin{eqnarray}
\dot{V}(\rho)=-\sum_{n=1}^kK T_n(t)^2\leq0.
\label{eqn:dvusl}
\end{eqnarray}
With such control fields, $V$ will decrease to its minimum  and
state $\rho$ will converge to the target state $\rho_f$ with the
same spectrum as the initial state $\rho_0$ and satisfying
$\Tr(e^{-iH_0t}\rho_fe^{iH_0t}[P,H_n])=0$ \cite{Alessandro}.

In the conventional  field design method, the amplitude of the control
fields $f_n(t)$ is proportional to $T_n$. That means when $T_n$ is
small (for example, when $\rho$ is very close to $\rho_f$), $f_n(t)$
will  become small leading to a slow decreasing of $V$ and a long
evolution time. Our aim is to determine optimized control fields
$f_n(t)$ to enable $V$ to decrease as fast as possible in order to speed
up control. From Eq.(\ref{eqn:dV}) it is seen that if each
$f_n(t)$ has a different sign to the corresponding $T_n$, then
$\dot{V}<0$ and large $|f_n(t)|$ will lead to fast decreasing of
$V$. Therefore, the problem has to be discussed under a
constraint on the control fields $f_n(t)$. Considering the following
reasons for constraining the control fields: First, one often wishes
control fields to be weak in order
to reduce energy (power) costs. Second, strong external control
fields may lead to invalidation of the modeling of the system.
Third, strong fields may disturb neighboring quantum systems that
we do not want to disturb.
We propose two designs of control fields under constraints on
the power of the control fields (constraint A)
and on the strength of each control field (constraint B), respectively.

\subsection{A power-type Constraint}
First, we consider the following  constraint on the control fields
\begin{eqnarray}
W=\sum_{n=1}^k f_n(t)^2\leq W_{max}.
\label{eqn:lmtA}
\end{eqnarray}
Since the control fields $f_n(t)$ are always associated with the
amplitude of the electro-magnetic fields, the quantity $W$ can be
interpreted as a power-type quantity. We will call it power for
simplicity in the following. The total power of the control fields is
bounded in this case.

Consider $\sum_{n=1}^k{T_n^2}\neq0$ ($\rho\neq\rho_f$) in  time $t$
and the constraint for $f_n(t)$ is $\sum_{n=1}^k f_n(t)^2=W$. In
order to determine the optimized control fields that minimize
$\dot{V}=\sum_{n=1}^kf_n(t)T_n$ ($\dot{V}$ is negative), we use the
Lagrange multiplier method. Let
\begin{eqnarray}
   L=\sum_n f_n T_n+\lambda (\sum_n f_n^2-W)
\label{eqn:Lag}
\end{eqnarray}
where $\lambda$ is the Lagrange multiplier, and $f_n$  represents
$f_n(t)$ at a certain time $t$. Then from the following equations
\begin{eqnarray}
\frac{\partial}{\partial f_n}L=T_n+2\lambda f_n=0,\\
\frac{\partial}{\partial \lambda}L=\sum{f_n^2}-W=0,
\label{eqn:dLag}
\end{eqnarray}
it is easy to obtain the amplitude of the fields $f_n$ at time $t$,
\begin{eqnarray}
f_n=-\frac{\sqrt{W}T_n}{\sqrt{\sum_{n=1}^k{T_n^2}}}
\label{eqn:fn}
\end{eqnarray}
with $T_n=\Tr(-i\rho[P,H_n])$. The corresponding time  derivative of
Lyapunov function is
\begin{eqnarray}
\dot{V}=\sum_{n=1}^kf_n T_n=-\frac{\sqrt{W}\sum_{n=1}^kT_n^2}{\sqrt{\sum_{n=1}^k{T_n^2}}}.
\label{eqn:dVA}
\end{eqnarray}
It is seen that $\dot{V}$ is proportional to $\sqrt{W}$, so we
choose $W=W_{max}$ for a faster decreasing of $V$ and our control
design for all evolution time reads
\begin{eqnarray}
f_n(t)=\left\{\begin{array}{c} -\frac{\sqrt{W_{max}}T_n}
{\sqrt{\sum_{n=1}^k{T_n^2}}}\ \ \ \ (\sum_{n=1}^k{T_n^2}\neq0) \\\ \
\ \ \ \ \ \ 0 \ \ \ \ \ \ \ \ \ \  (\sum_{n=1}^k{T_n^2}=0).
\end{array}
\right.
\label{eqn:fnA}
\end{eqnarray}
Note that when $\rho$ reaches the final state $\rho_f$, all
$T_n=\Tr(-i\rho_f[P,H_n])$ become zero and all control
fields are switched off. Considering that $\rho$ converges to $\rho_f$
asymptotically, we will switch off the control fields after
$D(\rho,\rho_f)<\varepsilon$ where $D$ denotes some measurement for
the distance between $\rho$ and $\rho_f$ and $\varepsilon$ is the
required precision.

In the case $k=1$, i.e., there is only one control Hamiltonian, our
control design reduces to
\begin{eqnarray}
f_1(t)=\left\{\begin{array}{c}
 \ -\sqrt{W_{max}}
 \ \ \ \ \ \ \ \ (T_1>0)\\
\ \ \  \sqrt{W_{max}}
 \ \ \ \ \ \ \ \ (T_1<0)\\
\ \ \ \ \ \ \ \  0
\ \ \ \ \ \ \ \ \ \ \ \ (T_1=0)
\end{array}
\right.
\label{eqn:A1}
\end{eqnarray}
which has a simple ``bang-bang" control form. With its discrete shape of the
control fields, this control design should be easy to realize
experimentally \cite{Yi,Zhou}.

\subsection{Constraint on the strength of each control field}
Next, we will find the optimized control fields when the  strength of
each field is bounded. For simplicity, we assume the maximum strength of
every control field $f_n(t)$ is $S$ $(S>0)$, i.e.,
\begin{eqnarray}
|f_n(t)|\leq S,\ \ \ \ (n=1,2,\cdots,k). \label{eqn:lmtB}
\end{eqnarray}
From $\dot{V}=\sum_{n=1}^kf_n(t)T_n$, it is easy to obtain  the
optimized control fields that minimize $\dot{V}$ with condition
Eq.(\ref{eqn:lmtB}),
\begin{eqnarray}
f_n(t)=\left\{\begin{array}{c}
\ -S\ \ \ \ \ \ \ \,(T_n>0)\\
\ \ \ S \ \ \ \ \ \ \ \ (T_n<0)\\
\ \ \ 0 \ \ \ \ \ \ \ \ \, (T_n=0)
\end{array}
\right. \ n=1,2,\cdots, k
\label{eqn:fnB}
\end{eqnarray}
and the time derivative of the Lyapunov function is
\begin{eqnarray}
\label{eqn:dVB}
\dot{V}=\sum_{n=1}^k f_n(t)T_n=-S\sum_{n=1}^k |T_n|.
\end{eqnarray}
This design has the ``bang-bang" control form with $k$ different
control fields. When there is only one control Hamiltonian, the
design has the same form as that in Eq.(\ref{eqn:A1}) with $\sqrt{W}$
replaced by $S$.

We have presented  two designs of control fields for systems
described by Eq.(\ref{eqn:Le}) and Lyapunov function
Eq.(\ref{eqn:Lya}) with two constraints. However, these designs can
also be applied to other Lyapunov control  as long as the derivative
of Lyapunov function has the form of Eq.(\ref{eqn:dV}) where
$f_n(t)$ represents a control field and $T_n$ represents a real
function of the quantum state. In fact, for many different kinds of
Lyapunov function and different dynamical equations
\cite{Grivopoulos,Cong,Wang2,Wang3}, $\dot{V}$ takes this form.

\section{illustrations}

In this section, we will present an example to illustrate the proposed
schemes. The example consists of a 3-level system driven by a
control Hamiltonian. We show that the system can be steered to an
eigenstate of the free Hamiltonian from arbitrary initial states
(except the states in the La Salle's invariant space) by both the
conventional design $f_n(t)=-KT_n$ and the design proposed in this
paper. The difference is that the present design can speed up the
convergence.

Consider a three-level system described by the quantum Liouville equation
\begin{eqnarray}
\frac{d\rho}{dt}=-i[H_0+\sum_{n=1}^{4}f_n(t)H_n,\rho]
\label{eqn:exmp}
\end{eqnarray}
with free Hamiltonian
\begin{eqnarray}
H_0=\omega\left(
       \begin{array}{ccc}
        1.5 & 0 & 0 \\
     0 & 1 & 0 \\
     0 & 0 & 0 \\
       \end{array}
     \right).
\label{eqn:H0}
\end{eqnarray}
\textbf{where the energy difference between $\ket{1}$ and $\ket{2}$
($\ket{2}$ and $\ket{3}$) is  $\omega$ ($\frac{1}{2}\omega$).
$\hbar=1$ has been set throughout this paper.}

The aim is to steer the system from an arbitrary  initial pure state
$\rho_0=\ket{\phi_0}\bra{\phi_0}$ to an eigenstate, say
$\ket{\phi_f}=\ket{3}=[1,0,0]^{T}$ of the free Hamiltonian $H_0$. We
choose the Lyapunov function $V=\Tr(P\rho)$ with
\begin{eqnarray}
P=\left(
       \begin{array}{ccc}
        0 & 0 & 0 \\
        0 & 1 & 0 \\
        0 & 0 & 1 \\
       \end{array}
        \right).
\end{eqnarray}
According to the Lyapuonv control theory, the system
Eq.(\ref{eqn:exmp}) will be driven to an eigenstate of $P$ with the
minimum eigenvalue $0$, i.e., $\rho\rightarrow\rho_f=\ket{3}\bra{3}$.
Additionally, in the pure state case, this Lyapunov function can be
explained as the Hilbert Schmidt distance between $\ket{\phi}$ and
$\ket{\phi_f}$. Recall  that
$P=\textrm{I}-\ket{\phi_f}\bra{\phi_f}$, and
$V=\Tr(P\rho)=\Tr((\textrm{I}-
\ket{\phi_f}\bra{\phi_f})\ket{\phi}\bra{\phi})=1-|\bra{\phi}\phi_f\rangle|^2,$
 our control design in this example acts to make the
distance between $\ket{\phi}$ and $\ket{\phi_f}$ decrease as fast as
possible with a given restriction on the control fields.

\begin{figure}
\includegraphics*[width=8.5cm]{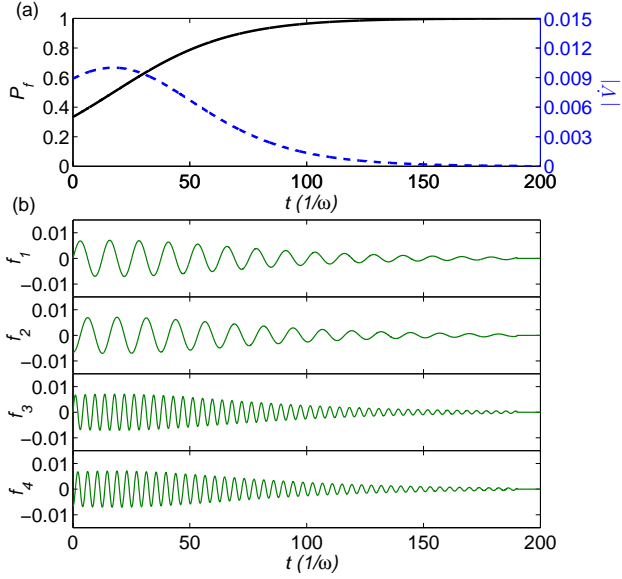}
\caption{\textbf{(color online)} (a) $P_f$ \textbf{(solid line)} and
$|\dot{V}|$ \textbf{(dashed line) }as a function of time with the
conventional  control design $f_n(t)=-KT_n$. (b) Control fields
\textbf{(in units of $\omega$) }for $W_{max}$=0.0001, and
$S_{max}=0.007$ . These fields are switched off when $P_f \geq
0.999$ at $t=191$ .} \label{FIG:usl}
\end{figure}

\begin{figure}
\includegraphics*[width=8.5cm]{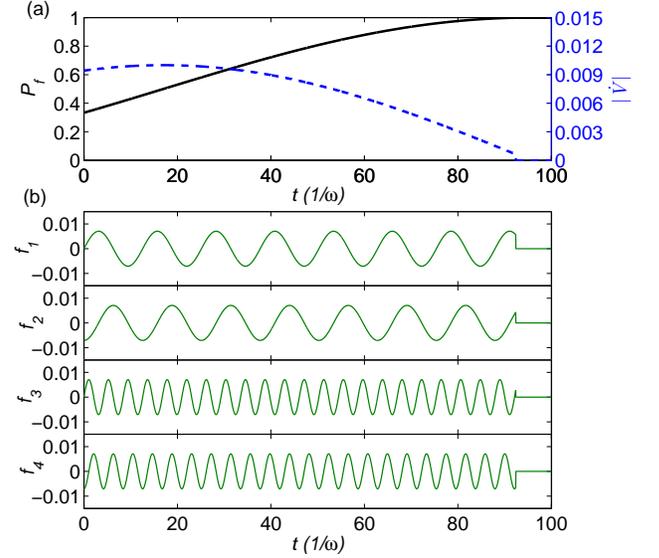}
\caption{\textbf{(color online)} (a) $P_f$ \textbf{(solid line)} and
$|\dot{V}|$ \textbf{(dashed line)} as a function of time with  the
control design Eq.(\ref{eqn:fnA}) under constraint A. (b) Control
fields \textbf{(in units of $\omega$)} with $W_{max}=0.0001$. These
fields are switched off at $t=90$ when $P_f \geq 0.999$.}
\label{FIG:optA}
\end{figure}

In order to achieve the best performance, we choose the
control Hamiltonians $H_n$ to be the generators of the $SU(3)$ group
$\lambda_n (n=1,...,8)$, namely, $H_c(t)$ is constrained from
 \begin{eqnarray}
\begin{split}
\lambda_1=\left(
       \begin{array}{ccc}
        0 & 1 & 0 \\
        1 & 0 & 0 \\
        0 & 0 & 0 \\
       \end{array}
     \right),
\ \ \ \ \ \ \
\lambda_2=\left(
       \begin{array}{ccc}
        0 & -i & 0 \\
        i & 0 & 0 \\
        0 & 0 & 0 \\
       \end{array}
     \right),
\\\lambda_3=\left(
       \begin{array}{ccc}
        1 & 0 & 0 \\
        0 & -1 & 0 \\
        0 & 0 & 0 \\
       \end{array}
     \right),
\ \ \ \ \ \ \
\lambda_4=\left(
       \begin{array}{ccc}
        0 & 0 & 1 \\
        0 & 0 & 0 \\
        1 & 0 & 0 \\
       \end{array}
     \right),
\\\lambda_5=\left(
       \begin{array}{ccc}
        0 & 0 & -i \\
        0 & 0 & 0 \\
        i & 0 & 0 \\
       \end{array}
     \right),
\ \ \ \ \ \ \
\lambda_6=\left(
       \begin{array}{ccc}
        0 & 0 & 0 \\
        0 & 0 & 1 \\
        0 & 1 & 0 \\
       \end{array}
     \right),
\\\lambda_7=\left(
       \begin{array}{ccc}
        0 & 0 & 0 \\
        0 & 0 & -i \\
        0 & i & 0 \\
       \end{array}
     \right),
\lambda_8=\frac{1}{\sqrt{3}}\left(
       \begin{array}{ccc}
        1 & 0 & 0 \\
        0 & 1 & 0 \\
        0 & 0 & -2 \\
       \end{array}
     \right).
\end{split}
\label{eqn:gen}
\end{eqnarray}
Notice that since only $\lambda_1,\lambda_2,\lambda_4$  and $\lambda_5$
satisfy $[P,\lambda_n]\neq0$, only these generators are effective
for our model which can be understood by examining
Eq.(\ref{eqn:dV}). Therefore, the control Hamiltonians are chosen
as,
\begin{eqnarray}
\begin{split}
H_1=\lambda_1=\left(
       \begin{array}{ccc}
        0 & 1 & 0 \\
        1 & 0 & 0 \\
        0 & 0 & 0 \\
       \end{array}
     \right),
     H_2=\lambda_2=\left(
       \begin{array}{ccc}
        0 & -i & 0 \\
        i & 0 & 0 \\
        0 & 0 & 0 \\
       \end{array}
     \right),\\
     H_3=\lambda_4=\left(
       \begin{array}{ccc}
        0 & 0 & 1 \\
        0 & 0 & 0 \\
        1 & 0 & 0 \\
       \end{array}
     \right),
     H_4=\lambda_5=\left(
       \begin{array}{ccc}
        0 & 0 & -i \\
        0 & 0 & 0 \\
        i & 0 & 0 \\
       \end{array}
     \right),
\end{split}
\end{eqnarray}
which can be rewritten  as  $H_1=\ket{3}\bra{2}+\ket{2}\bra{3}$,
 $H_2=i(-\ket{3}\bra{2}+\ket{2}\bra{3})$,
 $H_3=\ket{3}\bra{1}+\ket{1}\bra{3}$, and
 $H_4=i(-\ket{3}\bra{1}+\ket{1}\bra{3}),$ which couple
the energy levels $\ket{1}$ and $\ket{2}$ to the final state
$\ket{3}$. We would like to note that the control Hamiltonians  in
this example are optimal, because any operator for this 3-level
system can be written as an expansion of these generators. For
high-dimensional system, the problem becomes more complicated.

We first simulate the problem with the conventional  control  field
design $f_n(t)=-K\Tr(-i\rho [P,H_n])$ with $K=0.01$ and initial
state $\ket{\phi_0}=\frac{1}{\sqrt{3}}(\ket{1}+\ket{2}+\ket{3})$.
The control fields are switched off when the probability
$P_f=|\bra{\phi}\phi_f\rangle|^2$ reaches $0.999$ both in this
simulation and the following two so as to compare the evolution
time.  Fig.\ref{FIG:usl}(a) shows the evolution of  probability
$P_f$ (black solid line) and  $|\dot{V}|$ (blue dashed line). It is
seen that the system is driven to target state $\ket{\phi_f}$ and
the evolution time is about $t=191$ for $P_f=0.999$. The
time-varying control fields $f_n(t)$ are plotted in
Fig.\ref{FIG:usl}(b). The power $W=\sum_{n=1}^4f_n(t)^2$ reaches its
maximum $W_{max}=0.0001$ at $t=17$ and the maximal strength of a
single control field is $|f_1|=0.007$ at $t=16$.

\begin{figure}
\includegraphics*[width=8.5cm]{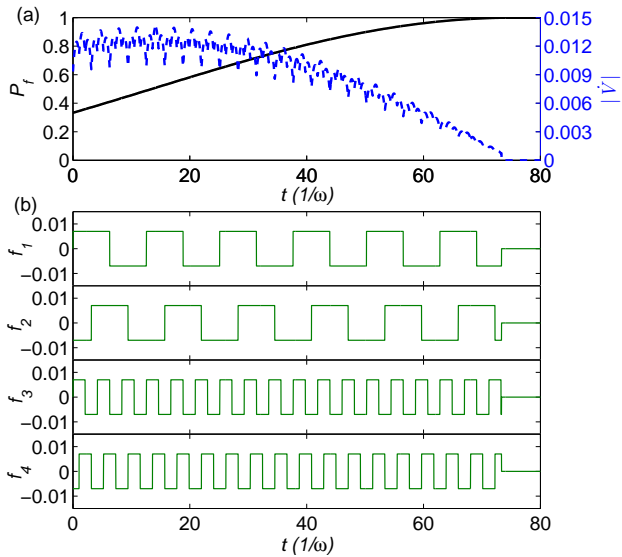}
\caption{\textbf{(color online) }(a) $P_f$ \textbf{(solid line)} and
$|\dot{V}|$ \textbf{(dashed line)} as a function of  time with
"bang-bang" control Eq.(\ref{eqn:fnB}) under constraint  B. (b)
Control fields \textbf{(in units of $\omega$ )} with $S=0.007$.
These fields are switched off at $t=74$ when $P_f \geq 0.999$.}
\label{FIG:bbc}
\end{figure}

Next, we employ the control design Eq.(\ref{eqn:fnA})  under the
constraint on the power of control fields. In order to be comparable
with the last example, the maximal power $W_{max}$ and initial state
are chosen to be the same as  the above one. Results in
Fig.\ref{FIG:optA}(a) show that the evolution time for $P_f=0.999$
is $t=92$ which is evidently shorter than that shown in Fig.
\ref{FIG:usl}. Control fields are plotted in Fig.\ref{FIG:optA}(b)
which have a sinusoidal shape. In this example, for all the time
before control fields are switched off, the power of control fields
remains constant and the shapes of fields are optimized so that the
Lyapunov function decreases fastest under constraint A. We can see
this from evolution of $|\dot{V}|$ in Fig.\ref{FIG:optA}(a) (blue
dotted line).

Now we study the control design Eq.(\ref{eqn:fnB})  under constraint
B. For comparison, we let the maximal strength $S$ and the initial
state be the same as that in the first simulation. The simulation
results are illustrated in Fig.\ref{FIG:bbc}. The evolution time for
$P_f=0.999$ is $t=73$ which is about $38$ percent of the first
example. In this example, the control fields shown in
Fig.\ref{FIG:bbc}(b) are step-like. Such a control method makes
Lyapunov function decrease fastest under the restriction of strength
and has the advantage of being easy to be implemented in experiment.
The evolution of $|\dot{V}|$ is not smooth (shown in
Fig.\ref{FIG:bbc}(a) by blue dashed line), which is due to the
discrete control fields.

\begin{figure}
\includegraphics*[width=8cm]{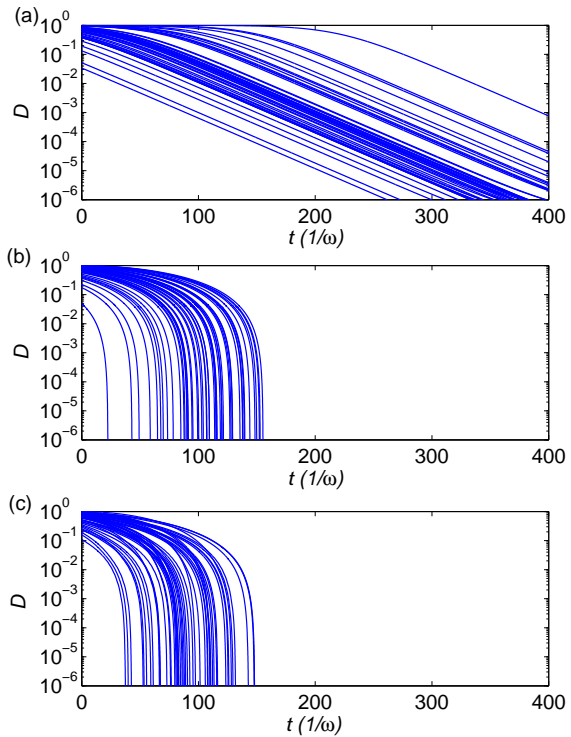}
\caption{(color online) Evolution of $D=1-|\bra{\phi}\phi_f\rangle|^2$  for the
three control designs  with logarithmic scale. (a),(b) and (c)
correspond to the conventional control field design, the design
under constraint A, and that under constraint B, respectively. Each
picture is a result averaged over 50 random initial states. The
convergence speed of (b) and (c) is faster than that of (a) as these
figures show.} \label{FIG:logcomp}
\end{figure}

Furthermore, we plot in Fig.\ref{FIG:logcomp} the  evolution of
$D=1-|\bra{\phi}\phi_f\rangle|^2$ for the  three designs of control
field with $50$ randomly chosen initial states ( $\ket{\phi_r}=R[r_1
e^{i2\pi r_4},r_2 e^{i2\pi r_5}, r_3 e^{i2\pi r_6}]^T$, where  $r_i
(i=1,...,6)$ are  random numbers uniformly created  between $0$ and
$1$, $R=\frac{1}{\sqrt{r^2_1+r^2_2+r^2_3}}$ is a normalization
factor) respectively. The simulations show that the convergence rate
for the usual control design is exponential where the evolution time
grows linearly with the distance between the actual and target states
$D$, whereas the convergence rate of our two methods is larger than
the conventional one,  especially when the control fidelity is
required to be high. The reason is our methods keep $|\dot{V}|$ at
its maximum for all the evolution even if $T_n$ is very small.

We found  that some of the control fields may oscillate with  very
high frequency at the end of the control in  Eq.(\ref{eqn:fnB}) with
constraint B. The oscillation depends on the initial state and the
distance between the actual and target states. The reason of this
oscillation is as follows. When the state is close to the target
state, some $T_n$ become very small leading to ineffectiveness of
the corresponding control field $f_n(t)$. While in conventional
Lyapunov  control, the control field $f_n(t)$ decreases to zero with
$T_n \rightarrow 0$. In  our method, however, $f_n(t)$ is designed
to take the value $S$ or $-S$, which makes the state oscillate
almost every step of simulation. This problem can be solved by
averaging the control fields over a proper time period  and then use
the reshaped fields instead of the oscillating one. In fact, this
average can be used in the situation when the control fields are not
a (fast) oscillating function of time, it yields the same control
fields and maintains the results.

It is worth noting that the convergency depends on the amplitude of
the control fields, when the control fields oscillate very fast, the
Lyapunov function $V$ almost stops to decrease. This is different
from the conventional Lyapunov design where $V$ keeps decreasing
with the amplitude of the control fields approaching zero. This
suggests that we can adjust the convergency by properly designing
the amplitude of the controls fields. We will show and prove  this
point  in the future work.

\section{Robustness of the designs}
One may wonder if these optimal designs improve the robustness of
the Lyapunov-based control. In the following, we shall examine this
problem  following  the representations in \cite{Yi2}  by
calculating an average  fidelity of the system with Hamiltonian
uncertainties, decoherence and  field fluctuation in the controls
Eq.(\ref{eqn:exmp}). Here the average  fidelity is defined by
$P^{'}_f=\frac 1 N(\sum_{j=1}^N|\bra{\phi_{j}}\phi_f\rangle|^2)$,
$|\phi_{j}\rangle$ denotes the actual  state evolving from a random
initial state under the control with uncertainties, decoherence or
fluctuations. In other words, the average is taken over $N$ actual
states, each evolves from a randomly chosen initial state, driving
by the controls with uncertainties, decoherence or fluctuations. Our
focus is on whether the optimal designs is robust against these
uncertainties compared with the conventional one.
\begin{figure}
\includegraphics*[width=8.5cm]{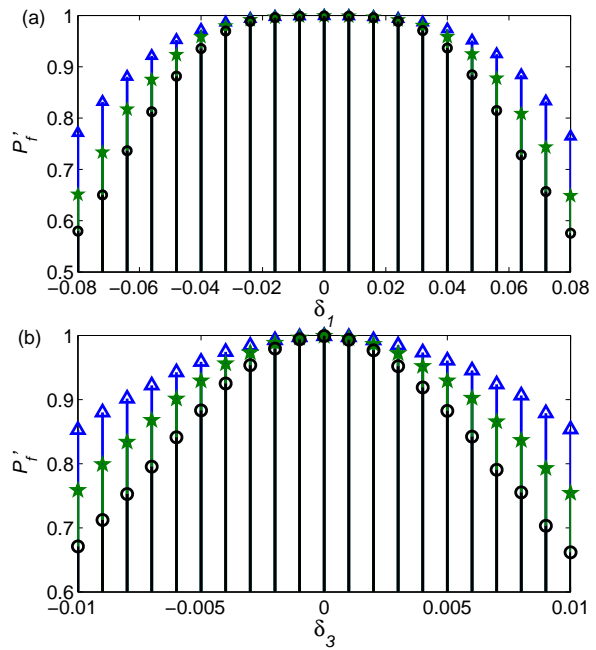}
\caption{\textbf{(color online)} The average fidelity  verses uncertainties in the free
Hamiltonian, (a) for  $\lambda_1$  and (b) for $\lambda_3$. The blue
triangle, green star and black circle represent the fidelity
obtained by bang-bang control design, design with power constraint
and the conventional one, respectively.} \label{FIG:hamrobust}
\end{figure}

We begin with analyzing robustness against the uncertainty in the
free Hamiltonian $H_0.$ The uncertainties can be taken into account
by adding a perturbation $\delta H_0$ to the free Hamiltonian, i.e.,
\begin{eqnarray*}
  H_0 \rightarrow H_0+\delta H_0.
\end{eqnarray*}
Here $\delta H_0=\sum_{n=1}^{8}\delta_n\lambda_n$ with $\delta_n$ a
real number and $\lambda_n$ the generators in Eq.(\ref{eqn:gen}).
For simplicity, we examine separately the 8 uncertainties $\delta
H_0 =\delta_n \lambda_n,$ $n=1,2\cdots 8$.  The equation
$\frac{d\rho}{dt}=-i[H_0+\delta H_0 +\sum_{n=1}^{4}f_n(t)H_n,\rho]$
is simulated for the three designs with the same control fields and
the same parameters as in section $\rm{III}$. Selected  results are
showed in Fig.\ref{FIG:hamrobust} where the fidelity $P^{'}_f$ is an
average over fidelities from 1000 randomly chosen  initial states.

\begin{figure}
\includegraphics*[width=8.5cm]{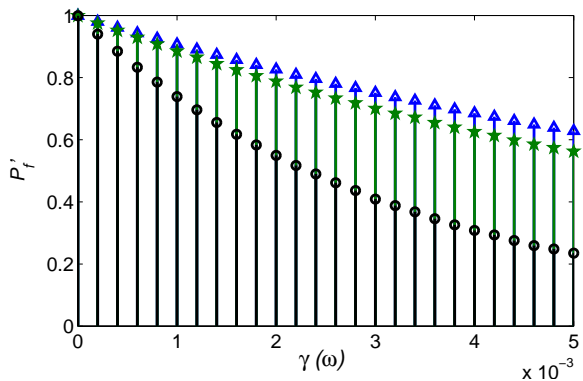}
\caption{\textbf{(color online)} The average fidelity  as a function of the decoherence
rates, here we choose $\gamma_2=\gamma_3=\gamma$. The blue triangle,
green star and black circle represent bang-bang control design,
design with power constraint and the conventional one,
respectively.} \label{FIG:Fid_gamma}
\end{figure}

The simulations show  that, (1) the optimal Lyapunov control is
robust against the uncertainty  $\lambda_1$, and is sensitive to
that of $\lambda_3$; (2) the bang-bang design is more robust than
the design with power constraint, and  the conventional design has
the worst robustness. In fact, the robustness against the
uncertainty $\delta_8\lambda_8$ is similar to
Fig.\ref{FIG:hamrobust}(a) and robustness against the uncertainties
$\delta_n\lambda_n (n=2,3,4,5,6,7)$ is roughly similar to
Fig.\ref{FIG:hamrobust}(b). Thus here we only show the robustness of
the control against the uncertainties $\lambda_1$ and $\lambda_3.$
This conclusion confirms that the system is sensitive to the
uncertainties that commute with the free Hamiltonian, while  it is
robust against  the uncertainties that does not.

We next discuss  the robustness against decoherence. In this
discussion, we assume that states $\ket{2}$ and  $\ket{1}$ are long
lived, namely  only the spontaneous emission
$\ket{3}\rightarrow\ket{1}$ and $\ket{3}\rightarrow\ket{2}$ is
assumed. This can be described by
$\mathcal{L}(\rho)=\sum_{i=1,2}\gamma_i(\sigma_i^-\rho\sigma_i^+
-\frac{1}{2}\sigma_i^-\rho\sigma_i^+-\frac{1}{2}\sigma_i^-
\rho\sigma_i^+)$ with $\sigma_i^-=\ket{i}\bra{3}$ and
$\sigma_i^+=\ket{3}\bra{i}$.  With this assumption,  the target
state is not a steady state, so the fidelity will be affected
seriously. However, the optimal controls are better than the
conventional one due to the short time  needed to drive the quantum
system from an initial state to  the target state. As we did in
Fig.\ref{FIG:hamrobust}, we calculate the average fidelity for the
three control designs with 1000 random initial state and fixed
$\gamma$. The results are depicted in Fig.\ref{FIG:Fid_gamma}. In
this case, the fidelity of the two optimal  designs  is obviously
higher than  the conventional design.

The robustness against field fluctuations (with zero-mean) and
errors in the initial state is also explored. The results are
similar to that in \cite{Yi2}, namely the control field fluctuation
with zero mean affects the fidelity slightly, while it depends
sharply on the fluctuation with non-zero mean, and the final
fidelity is sensitive to the initial state. In these cases, the
optimal control has no advantage with respect to the conventional
one.

\section{application to the cooling of a mechanical oscillator}
The cooling of mechanical resonators
\cite{Schilesser,Rae,xWang,Jacobs,Tian} becomes  an active research
topic  in recent years due to its potential applications in
detecting extremely small displacement and observing quantum
phenomenon of macroscopic mechanical object. In this section, we
apply the Lyapunov control to cooling a nano-mechanical resonator.
The results show that it is possible to cool a nano-mechanical
system to its ground state by Lyapunov control,  the optimized
control design leads to a shorter cooling time with respect to the
conventional control design.

Consider a nano-mechanical resonator (called target) with frequency
$\omega$ coupled to the other microwave (optical) oscillator
(auxiliary system),  the microwave oscillator  has a sufficiently
higher frequency $\Omega$ such that it can be prepared in its ground
state at finite  temperature. In the language of Lyapunov control,
the free  Hamiltonian of the composite system is given by
\begin{eqnarray}
H_0=\hbar\omega a^{\dag}a+\hbar\Omega b^{\dag}b.
\end{eqnarray}
We assume the coupling Hamiltonian of the two oscillators has the
following form,
\begin{eqnarray}
H_c=g(t)x_Ax_B
\end{eqnarray}
with $x_A=a+a^{\dag}$ and $x_B=b +b^{\dag}$. This type of
Hamiltonian can be realized by coupling the target to a LC
oscillator and the coupling rate $g(t)$ can be modulated by the
voltage of the LC circuit \cite{Tian,xWang,Jacobs}.

In the sideband cooling, $g(t)$ is modulated at  $\Omega-\omega$ so
that the two resonators are effectively coupled and the rotating
wave approximation (RWA) applies  when  the coupling $g$ is weak.
Recently, the authors of \cite{xWang} shown that quantum control can
improve the cooling, when the control goes beyond the RWA  in the
ultra-strong coupling regime $g \sim \omega$ \cite{xWang}. Here, we
show that we can obtain the control design by the Lyapunov
functional, and the optimized Lyapunov design can shorten the
cooling time.

Denote  the state of the two resonators  by $\rho$, we can choose
the Lyapunov function as
\begin{eqnarray}
V(\rho)=\Tr(a^\dag a \rho)=\langle n_a\rangle,
\end{eqnarray}
namely, we choose the mean phonon number  of the target resonator as
the Lyapunov function,  which is non-negative and becomes zero when
the target system is cooled to its ground state. By the same
procedure, we get
\begin{eqnarray}
\dot{V}(\rho)=g(t)T_n.
\end{eqnarray}
Here $T_n=\Tr(-i\rho[a^{\dag}a,(a^{\dag}+a)(b^{\dag}+b)])$. If
$g(t)$ is chosen to keep $\dot{V}\leq 0$,  the phonon number of
target resonator will decrease monotonically. The conventional
Lyapunov design for the time-dependent coupling strength is
$g(t)=-KT_n$ with $K$ a positive constant, while the optimal
Lyapunov design is a {\it bang-bang} control given by
Eq.(\ref{eqn:fnB}).

\begin{figure}
\includegraphics*[width=9.5cm]{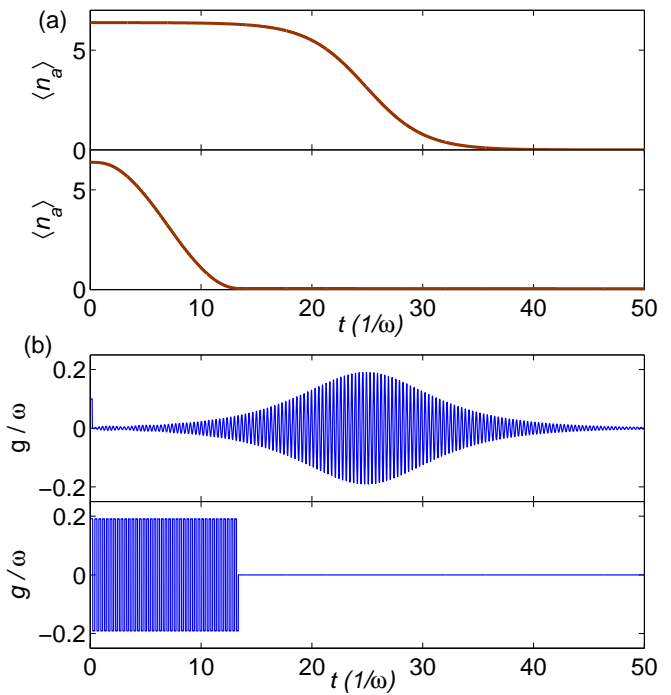}
\caption{(color online) (a) Phonon number versus time for the
conventional Lyapunov design (upper half)  and optimal Lyapunov
design (lower half). The evolution time for $\langle
n_a\rangle=0.05$ is $\omega t=37.5$ (upper) and $\omega t=13.4$
(lower), respectively.  (b) Time-dependent coupling for the
conventional design (upper) and the optimal design (lower), where
$g(t)$ for the two designs share the same maximal strength
$|g_{max}|/\omega=1.91$. $g(t)$ for the optimal design is shut off
after $\omega t=13.4$.} \label{FIG:col}
\end{figure}

Assume that  the target resonator is initially in a thermal state
with average phonon number $\langle n_a\rangle=6.38$ and the
auxiliary system is prepared in its ground state.  Note that the
frequency of auxiliary system $\Omega$ should be sufficiently large
compared with $\omega$, such that thermal fluctuation has small
effect on the microwave oscillator. Our numerical simulations show
that larger $\Omega$ leads to faster oscillation of $g(t)$, but it
does not affect the cooling results. Here we set
${\Omega}/{\omega}=20$.

The simulations are performed in the Fock space, we truncate the
Fock space of each oscillator up to 20 Fock states and the
dissipation of each resonator is ignored. Compare to simulations
with 25-Fock state-truncations, improvement is not significant, so
the simulations with 20-Fock state-truncation are reasonable. We
compare the evolution of phonon number $\langle n_a\rangle$ and
$g(t)$ for the two designs in Fig.\ref{FIG:col}. The top half of
Fig.\ref{FIG:col}(a) represents the evolution of $\langle
n_a\rangle$ for conventional design with $K=0.03$ and maximal
control field strength $|g(t)|/\omega=0.191$. It takes $t=37.5$ (in
unit of $1/\omega$) for the target to reach $\langle
n_a\rangle=0.05$. In contrast, for the optimal design, the evolution
time for reaching the same phonon number is $t=13.4$ as shown in the
lower half of Fig.\ref{FIG:col}(a). Obviously the optimal Lyapunov
design shorten the cooling time. Fig.\ref{FIG:col}(b) shows the time
dependence of $g(t)$ for the conventional (upper half) and optimal
(lower half) design.  For each case, the control field $g(t)$ starts
with a non-zero small number to avoid $T_n=0$ at the initial time.
It is seen that the optimal Lyapunov design leads to a faster
decrease of Lyapunov function (i.e. the phonon number). In addition,
the major components of the oscillation frequency of $g(t)$ is
automatically turned to the frequency difference $\Omega-\omega$ of
the two resonators(like that  in the sideband cooling scheme).

\section{summary}
We have presented two designs for  Lyapunov  control  under
constraints on the total power and individual strength of the
control fields. These designs make Lyapunov function decrease
fastest determined by the constraints. It has been shown that the
implementation of our designs leads to a shorter time towards the
target state especially for high fidelity requirement. Moreover, the
second control design gives  simple {\it bang-bang} control fields,
which may be easy to implement in experiment. Intuitively, our
methods use a constant power or strength of control fields to make
Lyapunov function decrease as fast as possible. This optimal control
is more robust against uncertainties in the Hamiltonian and
decoherence in the system with respect to the conventional design.
\textbf{We also explore the application of our optimal design to
cool a nanomechanical system,  a significantly shorter time is
obtained compared with the conventional Lyapunov design.} Here we
focused only on how to design control fields with fixed control
Hamiltonian $H_n$, a general formalism to choose control Hamiltonian
$H_n$ to speed up Lyapunov control is still an open issue.

This work is supported by NSF of China under grant Nos 61078011,
10935010 and 11175032, and the Australian Research Council.

\end{document}